\documentclass[sigconf]{acmart}

\usepackage{subcaption}
\usepackage{color,soul}
\usepackage{xcolor}
\AtBeginDocument{%
  }

\acmConference[Arxiv]{}{}{}

\begin{document}

\title{Posting Patterns of Members of Parental Subreddits}


\author{Nazanin Sabri}
\affiliation{%
  \institution{University of California, San Diego (UCSD)}
  \country{USA}}
\email{nsabri@ucsd.edu}

\author{Mai Elsherief}
\affiliation{%
  \institution{Northeastern University}
  \country{USA}}
\email{m.elsherif@northeastern.edu}

\renewcommand{\shortauthors}{Sabri et al.}

\begin{abstract}
Online forums (e.g., Reddit) are used by many parents to discuss their challenges, needs, and receive support. While studies have investigated the contents of posts made to popular parental subreddits revealing the family health concerns being expressed, little is known about parents' posting patterns or other issues they engage in. In this study, we explore the posting activity of users of 55 parental subreddits. Exploring posts made by these users (667K) across Reddit (34M posts) reveals that over 85\% of posters are not one-time users of Reddit and actively engage with the community. Studying cross-posting patterns also reveals the use of subreddits dedicated to other topics such as relationship and health advice (e.g., r/AskDocs, r/relationship\_advice) by this population. As a result, for a comprehensive understanding of the type of information posters share and seek, future work should investigate sub-communities outside of parental-specific ones. Finally, we expand the list of parental subreddits, compiling a total of 115 subreddits that could be utilized in future studies of parental concerns. 
\end{abstract}
\vspace{-5pt}

\ccsdesc[500]{Information systems~Social networks}
\ccsdesc[300]{Social and professional topics~User characteristics}

\keywords{Reddit, Parental Subreddits, Posting Pattern}

\received{10 December 2024}
\received[revised]{}
\received[accepted]{}

\maketitle

\vspace{-5pt}
\section{Introduction}

Many parents turn to online forums such as Reddit to share their concerns and ask for advice regarding their parental role and the well-being of their children~\cite{thornton2023scoping}. The affordances offered by such platforms (e.g., anonymity\footnote{For instance through the use of ``throwaway" accounts on Reddit.}) allow parents to discuss stigmatized issues such as postpartum depression, divorce, and custody~\cite{ammari2019self, mann2021emotional}. As parents often discuss intimate details of their parental experiences, priorities, and concerns, a large body of work has analyzed the contents of these postings to better understanding potential family health concerns~\cite{westrupp2022text, ossai2022using, stewart2016upvote, feldman2021because, francisco2024hey}. A number of these investigations of concerns were done through topic modeling~\cite{thornton2023scoping, westrupp2022text, o2021parenting}, reporting various concerns such as children's daily behaviors (e.g., food and eating patterns, sleep, and school), and conflicts of managing multiple-children~\cite{thornton2023scoping, westrupp2022text}. Prior work has also studied the experiences of specific subgroups such as parents of special education students~\cite{madsen2022communication} and foster families~\cite{lee2021using}. 

While the content of posts made by parents on Reddit has been studied, little attention has been paid to users' posting patterns such as user activity frequency, temporal engagement, or cross-community participation. Questions such as, whether posters to parental subreddits only utilize Reddit for their family health concerns or if they engage with other communities have not been answered. In this study, we answer the following research questions:

\begin{itemize}
    \item[RQ1:]What is the frequency and time span of posting to parental subreddits?
    
    \item[RQ2:]Do posters use subreddits other than those focused on the discussion of parental issues?  
\end{itemize}

\textbf{RQ1} explores how users engage with online parenting communities, examining the duration and frequency of engagement sheds light on whether these communities serve as long-term support systems or temporary coping mechanisms. \textbf{RQ2} addresses whether users compartmentalize their Reddit activity to parenting-related issues or engage with broader communities. Investigating cross-community participation helps reveal whether parental subreddit users are primarily concerned with parenting topics or integrate their Reddit activity with other personal or professional interests~\cite{singer2014evolution}.
These insights into usage patterns of Reddit have implications for developing strategies to better engage and support cross-community collaboration between parenting subreddits and other relevant communities 
to address multifaceted challenges faced by parents.

\vspace{-8pt}
\section{Data}

To study posting behavior on parental subreddits, we first collected Reddit posts from June 2005 to December 2023~\cite{reddit_data_source}. Over 2.4 billion posts were collected from this source, posted in the aforementioned period of 19 years. We then removed posts authored by ``AutoModerator", or written by users who had later deleted their account (i.e., ``[deleted]"). We excluded these users from our analysis since they cannot be tracked across sub-communities. We then used existing lists of parental subreddits from prior work~\cite{10.1145/3411764.3445203}, as well as searched on Reddit and the web to curate a list of 55 subreddits that are dedicated to the discussion of parental issues on Reddit. The full list is displayed in Table \ref{table:list_of_subreddits}. 

\begin{table*}
    \centering
    \resizebox{1.95\columnwidth}{!}{
    \begin{tabular}{p{2cm}|p{15cm}}
        \multicolumn{2}{c}{\textbf{List of Parenting Subreddits}} \\\hline\hline 
         Original List (N=55)&r/Parenting, r/AttachmentParenting, r/breakingmom, r/ParentingADHD, r/beyondthebump, r/raisingkids, r/Mommit, r/ParentingLite, r/ParentingTech, r/TeenParenting, r/stepparents, r/NewParents, r/ZeroWasteParenting, r/parentingteenagers, r/ParentingInBulk, r/MuslimParenting, r/CatholicParenting, r/AsianParentStories, r/LGBTParenting, \colorbox{yellow}{r/lgbtfamilies}, r/cisparenttranskid, r/Adoption, r/regretfulparents, \colorbox{yellow}{r/Singlemomsupport}, r/workingmoms, r/BabyBumps, r/breastfeeding, r/breastfeedingsupport, r/Buyingforbaby, r/FormulaFeeders, r/NewMomStuff, r/pottytraining, r/SingleParents, r/SAHP, r/clothdiaps, r/downsyndrome, r/homeschool, r/NewDads, r/predaddit, \colorbox{yellow}{r/radicalparenting}, r/SpecialNeedsChildren, r/UKParenting, r/Parents, r/BabyExchange, r/boobsandbottles, r/BreastPumps, r/NaturalPregnancy, r/parent, r/teachingchildren, r/birthcontrol, \colorbox{yellow}{r/parentdeals}, r/playgroup, r/TryingForABaby, r/FamilyDinner, r/pregnant\\\hline 
         Amendments (N=60)& r/americandad, r/daddit, r/AskParents, r/Midwives, r/parentsofmultiples, r/thingsmykidsaid, r/cutekids, r/vbac, r/toddlers, r/Preterms, r/atheistparents, r/Boymom, r/Britishdads, r/Breakingdads, r/dad, r/dadcore, r/calgarymoms, r/beyondbaby, r/bigbabiesandkids, r/babywearing, r/babywearingmoms, r/VelcroBabies, r/babystroller, r/BabyStrollers, r/babystuffthatyouneed, r/babyrooms, r/babyprep, r/babyprints, r/stepkids, r/splitparenting, r/stayathomemoms, r/Singlefather\_bychoice, r/Stepdadreflexes, r/Stepmom, r/stepmomreflexes, r/altmoms, r/animalstoriesforkids, r/babyloss, r/trollingforababy, r/amipregnant, r/baby,  r/dadswhodidnotwantpets, r/firsttimemom, r/happycryingdads, r/family, r/kidscrafts, r/kidsfallingdown, r/kidsfun, r/kidsfunnydrawings, r/KidsFunLearning, r/mombloggers, r/DadBloggers, r/momtokgossip, r/mothersday, r/FathersDay, r/FathersDayGifts, r/MothersDayGiftIdeas, r/singlemoms, r/weightlossafterbaby, r/healthypregnancy\\\hline 
    \end{tabular}
    }
    \caption{List of subreddits dedicated to the discussion of parenting. The original list was used to create the dataset used in this work. The amendments are new subreddits identified through our data analysis that could be used for discussion of parenting, pregnancy, or children.}
    \label{table:list_of_subreddits}
    \vspace{-25pt}
\end{table*}

Having identified these 55 parental subreddits, we filtered posts to those made in one of these subreddits (N = 2,189,656). We then extracted the usernames of users who had made these posts, finding a total of 667,825 unique usernames. Finally, we iterated the full dataset again and extracted all posts made by these users. This round of filtering resulted in compiling a total of 44,022,667 posts (2,194,480 are posts made to parenting subreddits, and 41.8M to other subreddits). These posts were made in 390,728 unique subreddits. 

\begin{figure*}[t!]
    \centering
    \begin{subfigure}[t]{0.31\textwidth}
        \centering
        \includegraphics[height=1.4in]{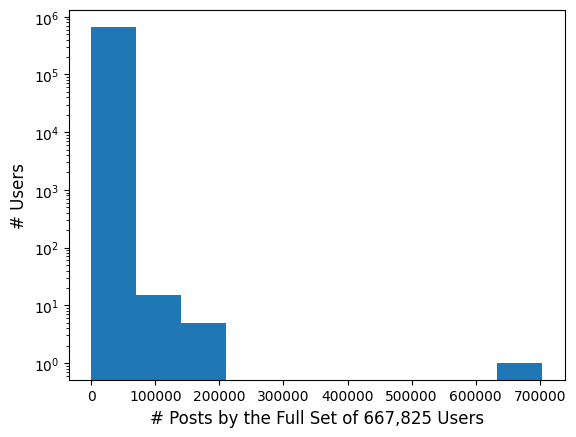}
        \caption{Histogram of number of posts per user for the full set of users (Log Scale).}
        \label{figure:full_users_activity}
    \end{subfigure}\hfill
    \begin{subfigure}[t]{0.31\textwidth}
        \centering
        \includegraphics[height=1.4in]{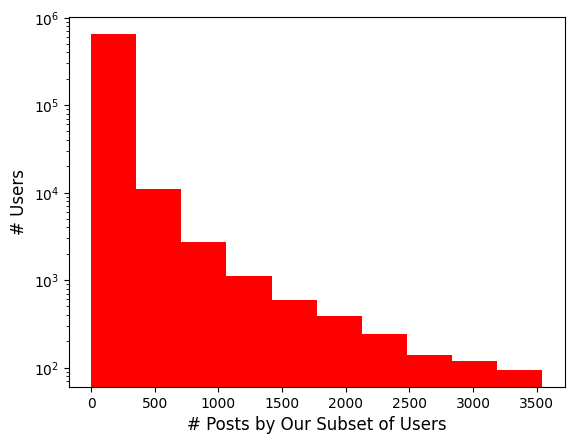}
        \caption{Histogram of number of posts per user after removing users with anomalous activity (Log Scale).}
        \label{figure:subset_users_activity}
    \end{subfigure}\hfill
    \begin{subfigure}[t]{0.31\textwidth}
        \centering
        \includegraphics[height=1.4in]{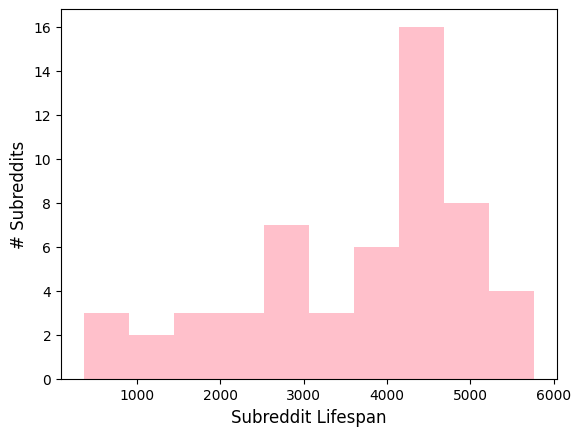}
        \caption{Histogram of the lifespan of parental subreddits in days.}
        \label{figure:parental_subreddit_lifespan}
    \end{subfigure}\\
    \begin{subfigure}[t]{0.31\textwidth}
        \centering
        \includegraphics[height=1.4in]{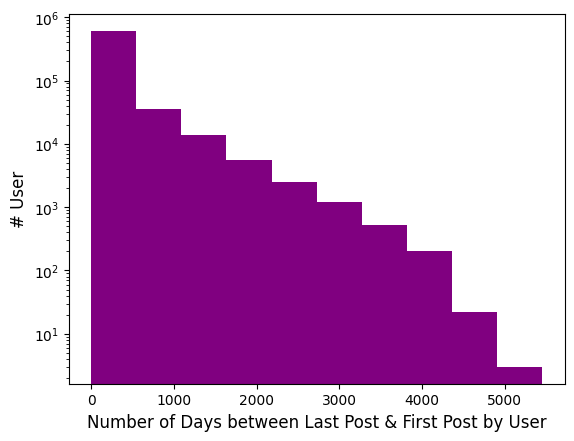}
        \caption{Histogram of the duration of account activity (last - first post) within parental subreddits (Log Scale).}
        \label{figure:user_lifespan}
    \end{subfigure}\hfill
    \begin{subfigure}[t]{0.31\textwidth}
        \centering
        \includegraphics[height=1.4in]{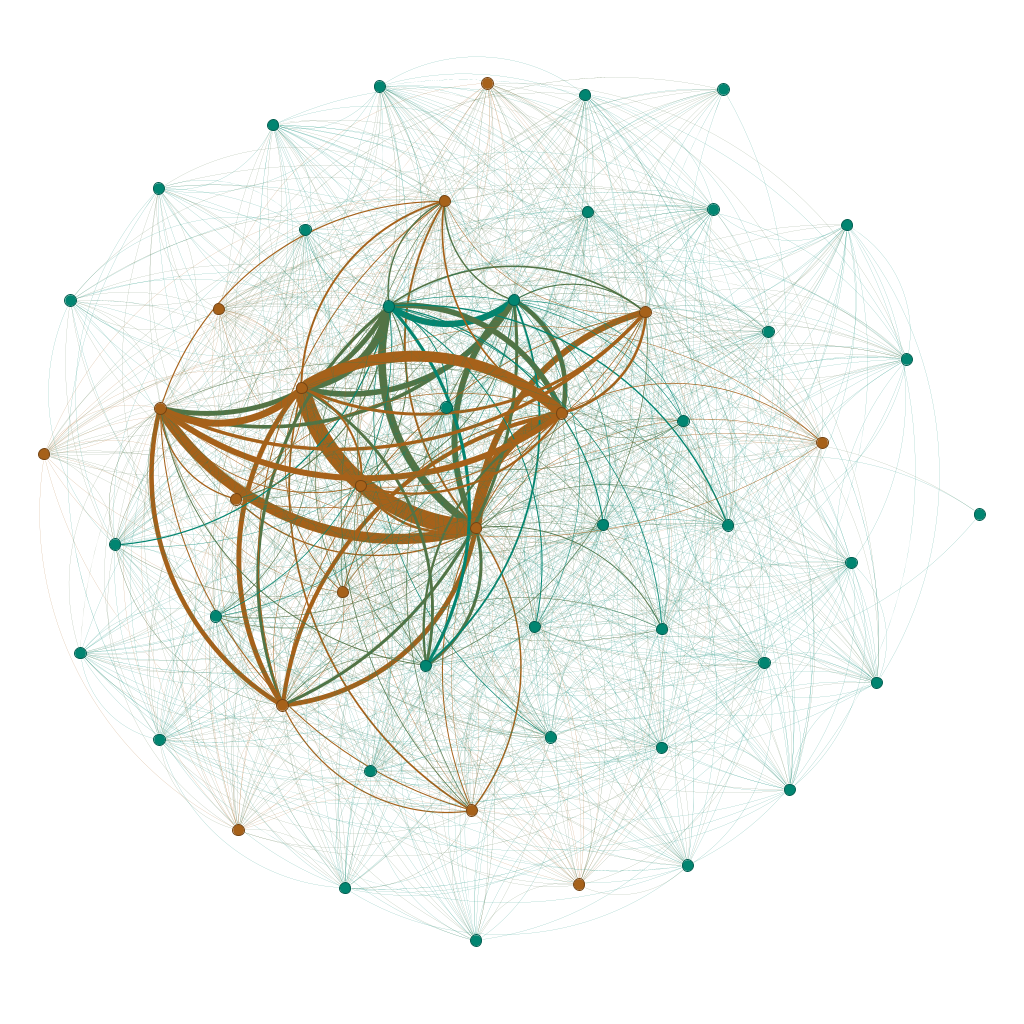}
        \caption{Undirected weighted graph of connections between parental subreddits ($Min(w_e) = 1$). Nodes are colored based on the assigned modularity class.}
        \label{figure:graph_parental_non_directed}
    \end{subfigure}\hfill
    \begin{subfigure}[t]{0.31\textwidth}
        \centering
        \includegraphics[height=1.4in]{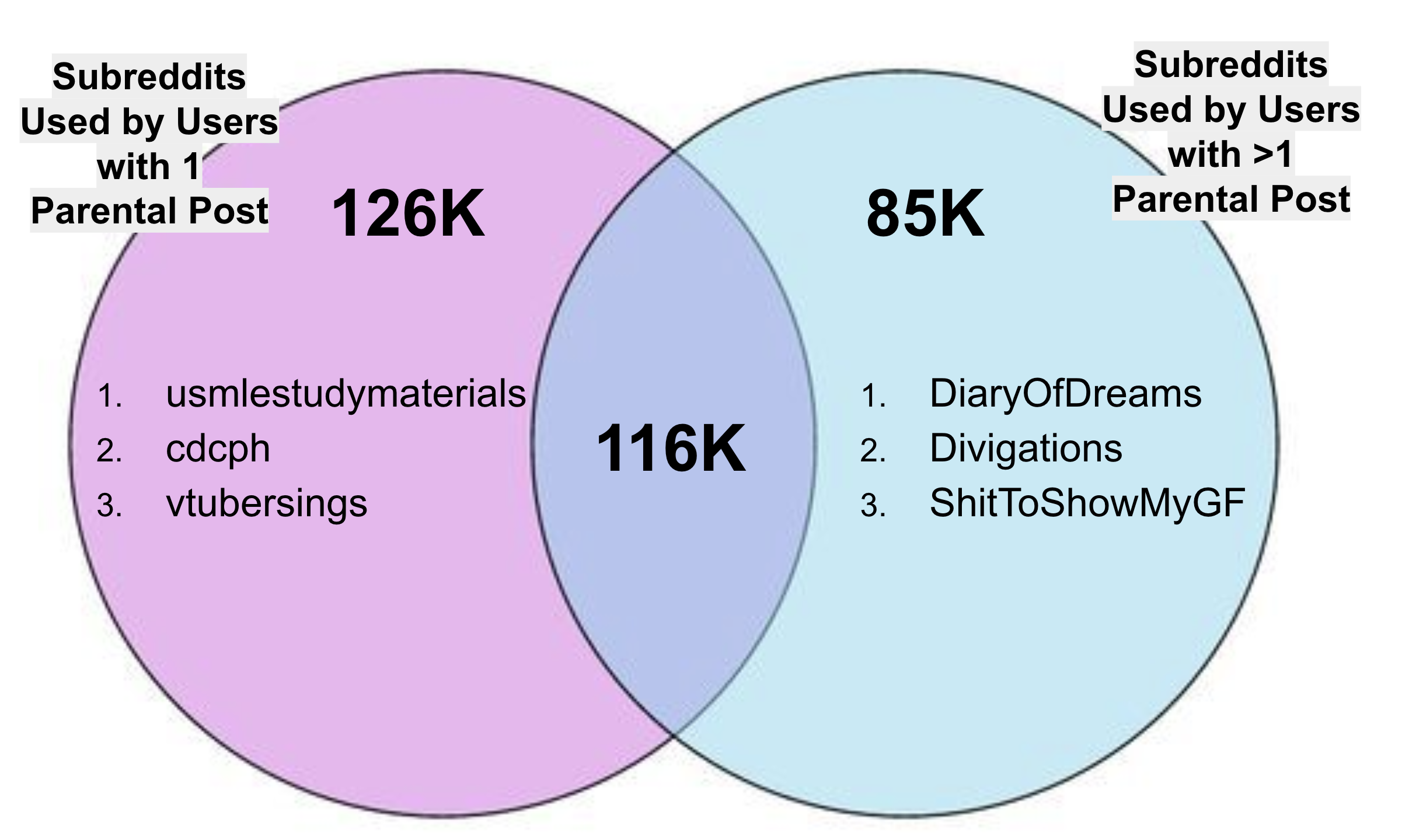}
        \caption{Venn diagram of the number of subreddits one-time and frequent users of parental subreddits use. The 3 subreddits mentioned are the most used within the subgroups.}
        \label{figure:venn_diagram}
    \end{subfigure}
    \caption{Visualizations of activity patterns.}
    \vspace{-5pt}
\end{figure*}

To better understand our dataset of 44,022,667 posts, we visualize the number of posts per user displayed in Figure \ref{figure:full_users_activity}. We can see that a small number of users had created significant number of posts (maximum number of posts by a single user = 703,500). However, the majority of users had much fewer posts, with the average number of posts per user being 65.92 (median = 13, standard deviation = 1,161.32, and minimum = 1). We decided to remove all activity of any user who had created more than $mean + 3 * STD = 3,549.89$ posts as values beyond this point can be considered to be anomalies~\cite{anomaly_formula,rousseeuw1986robust}. This threshold excludes 684 users from our analysis, reducing the number of unique users within our dataset from 667,825 to 667,141. The distribution of user activity by the subset of users we keep is shown in Figure \ref{figure:subset_users_activity}. The number of posts we utilize within our analysis is also reduced from 44,022,667 to 34,766,546.

\vspace{-5pt}
\section{Method}
\label{section:method}

To study cross-community posting patterns by users of parental subreddits we construct weighted graphs. In the graphs we create, each node represents a subreddit ($S_i$). Subreddits are connected to one-another using weighted edges. The weight of the edge ($w_{ij}$) represents the unique number of users who had posted to both $S_i$ and $S_j$. We create this graph twice, once between the original 55 parental subreddits, and once including all subreddits. We discuss the properties of these graphs in the results section. 

\vspace{-5pt}
\section{Results}

We begin our analysis by focusing on posts made within parental subreddits (N = 2,189,656). Statistics of posts per user and subreddit are displayed in Table \ref{tab:statistics_of_posting}. Figure \ref{figure:parental_subreddit_lifespan} displays the lifespan of the 55 subreddits we analyzed. We define subreddit lifespan as the number of days between the first and last post made to the subreddit between 2005-2023. The minimum amount of time these subreddits are active for is 357 days (less than a year) with the majority being active for around 10-13 years. We also look at how long users were active exclusively within parental subreddits, displayed in Figure \ref{figure:user_lifespan}. We can see that a large number of users were active for 0 days. This is because a large number made a total of 1 post in parental subreddits. We will further analyze this subset of users shortly. However, we can also see that a considerable portion of users (40\%) kept their accounts active for far longer (up to over 13 years).


\begin{table}[]
    \centering
    \begin{tabular}{|l|c|c|c|c|}
        \hline 
        \multicolumn{1}{|c|}{Activity} & \multicolumn{2}{c|}{Parental} & \multicolumn{2}{c|}{Non-Parental}\\
        \cline{2-5}
        \multicolumn{1}{|c|}{Statistics} & U & SR & U & SR \\
        \hline
        \# & 667K & 55 & 571K & 328K \\ \hline
        Minimum   & 1 & 4 & 1 & 1 \\
        Mean &   3.28  & 39,811.93 & 57.04 & 99.02 \\
        Median & 1 & 3,482 & 15 & 2 \\
        Maximum & 3,003 & 398,575 & 3,541 & 985,454 \\
        \hline
    \end{tabular}
    \caption{This table displays statistics regarding posts per user (column $U$) and per subreddit (column $SR$). The set of columns labeled as ``Parental" displays these statistics within the 55 subreddits identified as dedicated to the discussion of parental issues. ``Non-Parental" columns show these values in other subreddits within our dataset.}
    \label{tab:statistics_of_posting}
    \vspace{-25pt}
\end{table}

To better understand cross-posting patterns within parental subreddits, we create the parental subreddit connection graph (Section \ref{section:method}). This graph, displayed in Figure \ref{figure:graph_parental_non_directed}, is made up of 53 nodes, and 1,111 edges. We can see that the number of nodes (53) is less than our initial set of parental subreddits (55). \textit{r/teachingchildren} and \textit{r/LGBTParenting}, with a total of 66 and 4 posts respectively, have no shared users with other parental subreddits. Within this graph, the minimum edge weight is 1, and the maximum is 26,912 (average = 390). The edge with maximum weight is between \textit{r/BabyBumps} and \textit{r/beyondthebump} indicating that 26,912 of our users posted to both of these subreddits. Calculating modularity (0.052), we find that subreddits are grouped into 2 communities. A manual inspection of subreddits categorized within each of the two communities reveals that subreddits about pregnancy and new parents are placed in one class, and the remainder within a separate class. We also find the average path length within the graph to be 1.19 (diameter = 2). These findings imply that the majority of parental subreddits are very closely connected, with users posting across multiple subreddits. If we place a limit of above 10 shared users between two subreddits, our graph is reduced to 604 edges and 49 nodes. The subreddits removed due to this limit have been highlighted yellow in Table \ref{table:list_of_subreddits}. This graph is still highly connected with an average Clustering Coefficient of 0.874.

As discussed, a large number of users (60\%, N = 403K) had made only 1 post within parental subreddits. Tracking other postings by this group of users, we find that 19.5\% (N = 78K) did not make any other posts. The other 325K users with one submission within parental subreddits, had made a total of 16,061,051 other posts. We include the top 10 non-parental subreddits utilized by these users in Table \ref{tab:top_non_parental_subreddits}. We can see that the activity of these users is largely similar to those who post more often in parental subreddits. Members of parental subreddits are shown to be frequent members of other lighthearted subreddits (e.g., r/funny), as well as subreddits that don't stigmatize asking questions (e.g., r/NoStupidQuestions). We also observe the use of subreddits that could offer advice on relationships (e.g., r/shittyrelationships, r/relationshipproblems), health (e.g., r/healthcare, r/Dietandhealth), or finance (e.g., r/personalfinance, r/povertyfinance) by this population. Figure \ref{figure:venn_diagram} displays the difference between subreddits used by one-time users of parental subreddits compared to more frequent users.

\begin{table}[]
    \centering
    \begin{tabular}{|p{3.8cm}|p{4cm}|}
        \hline 
        Users with 1 post in Parental & Users with >1 post in Parental \\
        \hline\hline
        r/AskReddit	(483K) & r/AskReddit (502K)\\
        r/funny	(129K) & r/funny (122K)\\
        r/pics	(102K) & r/aww (117K)\\
        r/aww	(93K) & r/pics (110K)\\
        r/Showerthoughts	(77K) & r/cats (76K)\\
        r/politics	(71K) & r/NoStupidQuestions (73K)\\
        r/videos	(68K) & r/Showerthoughts (72K)\\
        r/NoStupidQuestions	(65K) & r/politics (60K)\\
        r/cats	(56K) & r/videos (56K)\\
        r/relationship\_advice	(55K) & r/AskDocs (55K)\\
        \hline
    \end{tabular}
    \caption{Top non-parental subreddits used by users of parental subreddits. These top non-parental subreddits are separated by those used by users who only posted once to parental subreddits and those who posted more to parental subreddits. The number shown in parenthesis is the number of posts made by that subset of users within each subreddit.}
    \label{tab:top_non_parental_subreddits}
    \vspace{-25pt}
\end{table}

To better understand cross-posting behaviors outside of parental subreddits, we create a graph of connections between all subdreddits included in our dataset. Out of the 328K subreddits used by all users of parental subreddits (including the parenting subreddits themselves), 184K (55.9\%) were only used by a single unique user (who could have made one or a lot of posts). Only 14.8\% (48K) had more than 10 unique posters and 13,427 (4\%) more than 100 unique posters. Given the large number of users within our dataset (N = 667K) we focus on the subreddits used by more than 100 unique users. We first use these 13K subreddits within our dataset to find other parental subreddits. To find these subreddits, we first find any that include parent or child keywords (i.e., mom, mother, dad, father, baby, kid, and pregnant). This search returns 55 subreddits, which we then manually inspect. The set of parental subreddits among these is displayed in Table \ref{table:list_of_subreddits} in the \textit{Amendments} row. While performing these manual searches on Reddit, a number of other suggestions came up which were also included. We believe the total list of 115 subreddits can help future work inspect aspects of parenting, pregnancy, and family issues from more diverse perspectives and communities. 



\begin{figure}
    \centering
    \includegraphics[width=0.38\linewidth]{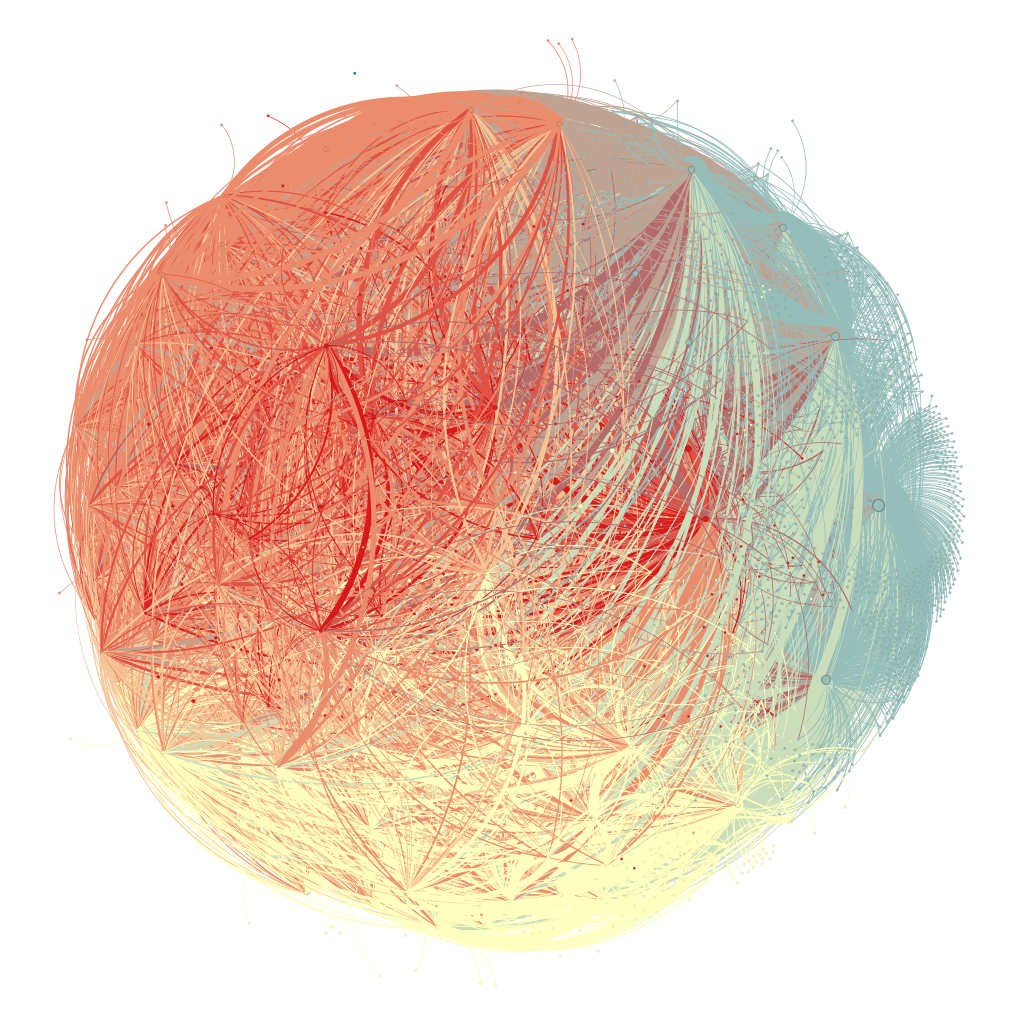}  
    \caption{Graph of 2K subreddits connected based on shared posters. Nodes are colored based on modularity class.}
    \label{fig:all_subreddits}
    \vspace{-15pt}
\end{figure}

Finally, we use the set of 13K subreddits with more than 100 unique users in our dataset to create a subreddit connection graph (described in Section \ref{section:method}). This graph has 13,427 nodes and 49,498,736 edges. The weights of these edges (representing the number of shared users) have a minimum value of 1, $average = 7.9$, and $median = 3$. The two subreddits with the most number of shared users are \textit{r/AskReddit} and \textit{Parenting} with $w=50K$. To be able to visualize this graph (shown in Figure \ref{fig:all_subreddits}), we limit edges to those with $w_e > 100$ (more than 100 users posting to both subreddits). This reduces the graph to 2,014 nodes and 31,832 edges. The graph is made up of two weakly connected components. Modularity analysis on these 2K subreddits creates 5 categories. While most categories are too broad to be assigned a theme, one only includes subreddits dedicated to ``onlyfans". We also use the \textit{Statistical Inference} feature of Gephi to find 108 inferred classes. The topics of some of these classes (identified through manual inspection) are music, finance, and different social media platforms. 

\vspace{-6pt}
\section{Limitations \& Future Work}

In this work we investigated the posting behavior of users of parental subreddits. We focused our analysis on posts and did not investigate comments that were being made by users. In our future work we aim to investigate users engagement with different Reddit sub-communities through other means such as commenting. Additionally, in this work we did not investigate the content of posts and how the contents change through time. Future work could look into how the contents of user's posts change in time, and if these changes correlate with their children's age and growth. Finally, it is important to note that prior work has demonstrated that Redditors utilize ``throwaway" accounts to preserve their identity when discussing stigmatized topics (e.g., parenting issues)~\cite{ammari2019self}. As a result, our work, and similar investigations could not account of users of the platform who choose not to post under a persistent username. 

\vspace{-4pt}
\section{Conclusion}

In this work we analyzed the posting patterns of users who had posted to 55 parenting subreddits at least once. While 60\% of users only made one post in the subset of 55 parental subreddits, the majority of these one-timers, posted to other subreddits. We also found that many users kept their accounts active for over a year and up to 13 years. In summary our findings 
challenge the assumption that contributors to parental subreddits are relatively inactive, primarily posting to address specific concerns. Instead, we discovered that this population is made up of highly active Reddit users who participate across a diverse range of subreddits covering a variety of topics. 

\vspace{-5pt}

\bibliographystyle{ACM-Reference-Format}
\bibliography{sample-base}


\end{document}